\documentclass[twocolumn,showpacs,prl,aps]{revtex4}
\usepackage{amsmath}
\usepackage{amsfonts}
\usepackage{amssymb}
\usepackage{graphicx}
\usepackage{hyperref}
\usepackage{bm}
\usepackage{color}
\hypersetup{backref,
colorlinks=true,
linkcolor=cyan,
linktoc=page,
citecolor=cyan}

\newcommand{\pr}[1]{\ensuremath{\left[#1\right]}} 
\newcommand{\pc}[1]{\ensuremath{\left(#1\right)}} 
\newcommand{\px}[1]{\ensuremath{\left\lbrace#1\right\rbrace}} 
\newcommand{\bra}[1]{\ensuremath{\left\langle#1\right\vert}} 
\newcommand{\ket}[1]{\ensuremath{\left\vert#1\right\rangle}} 
\newcommand{\md}[1]{\ensuremath{\left\vert#1\right\vert}} 
\newcommand{\av}[1]{\ensuremath{\left\langle#1\right\rangle}} 

\DeclareMathOperator{\Tr}{Tr}

\begin{document}

\title{Sympathetic laser-cooling of graphene with Casimir-Polder forces}
\author{Sofia Ribeiro}
\email{sofia.ribeiro@lx.it.pt}
\author{Hugo Ter\c{c}as}
\affiliation{Instituto de Telecomunica\c{c}\~oes, Lisbon, Portugal}


\begin{abstract}
We propose a scheme to actively cool the fundamental flexural (out-of-plane) mode of a graphene sheet via vacuum forces. Our setup consists in a cold atom cloud placed close to a graphene sheet at distances of a few micrometers. The atoms couple to the graphene membrane via Casimir-Polder forces. By deriving a self-consistent set of equations governing the dynamics of the atomic gas and the flexural modes of the graphene, we show to be possible to cool graphene from room temperatures by actively (laser) cooling an atomic gas. By choosing the right set of experimental parameter we are able to cool a graphene sheet down to $\sim 60~\mu$K.
\end{abstract}
\maketitle

A lot of attention has been drawn in the last years to the development of quantum technologies with hybrid quantum systems whose elementary building blocks are of different nature \cite{RMP85_2313-2363_2010}. The general trend is to combine well-characterized individual quantum systems, such as trapped ions \cite{tian, schmid}, degenerate quantum gases \cite{proukakis}, superfluid or superconducting Josephson junctions \cite{grupp}, quantum dots \cite{recati}, and nanomechanical oscillators \cite{chan, poot}, with microwave guides, optical resonators, and fibers \cite{mekhov}. The most promising applications of such systems range from high-precision force and mass measurements to quantum computation \cite{gavartin, machnes, schneider, hunger}. Optomechanical setups have been particularly successful in that task, making possible to cool down a mechanical system to its quantum ground state \cite{marquardt_review}. Radiation-pressure cooling of nano- or micromechanical cantilevers \cite{metzger, gigan, arcizet, kleckner, regal}, vibrating microtoroids \cite{carmon, schliesser}, and membranes \cite{thompson} constitute important hallmarks in the field of optomechanics with important implications in quantum technologies. 

With the advent of graphene and other two-dimensional materials, the zero-point cooling of macroscopic membranes becomes an imperative to the development of quantum technology based on suspended graphene. At low temperatures, the electrical resistivity in graphene is essentially hindered by the scattering between the electrons and the flexural (out-of-plane) phonons \cite{viljas, laitinen}, which one could, in principle, cool down with the help of an optomechanical setup. However, given the broad-band optical transmission of graphene (a graphene mirror is typically 98\% transparent) \cite{stauber}, the coupling between the macroscopic mechanical motion and the photons is very unlikely, making radiation-pressure cooling totally ineffective \cite{barton}. As such, dilution refrigerators have been used in the attempt to approach the quantum limit. Recently, narrow-gap microwave-cavity cooling of graphene has brought the thermal motion of graphene down to $\sim$60~mK \cite{PRL113_027404_2014}. It is therefore natural to investigate suitable alternatives to radiation-pressure schemes by exploiting the advantage of interfaces with cold atoms. 
Recent results showed that sympathetic cooling with ultracold atoms enables to reach ultralow temperatures in levitated optomechanical systems originally at room temperature when direct laser or evaporative cooling is not possible \cite{PRA91_013416_2015}. Also, it as been showed that heat transfer between two parallel layers of dipolar ultracold Fermi gases at different temperatures via dipolar couplings could be used this as an effective cooling process \cite{arXiv:1601.05949}.
\begin{figure}[t!]
\centering
  \includegraphics[width=8cm]{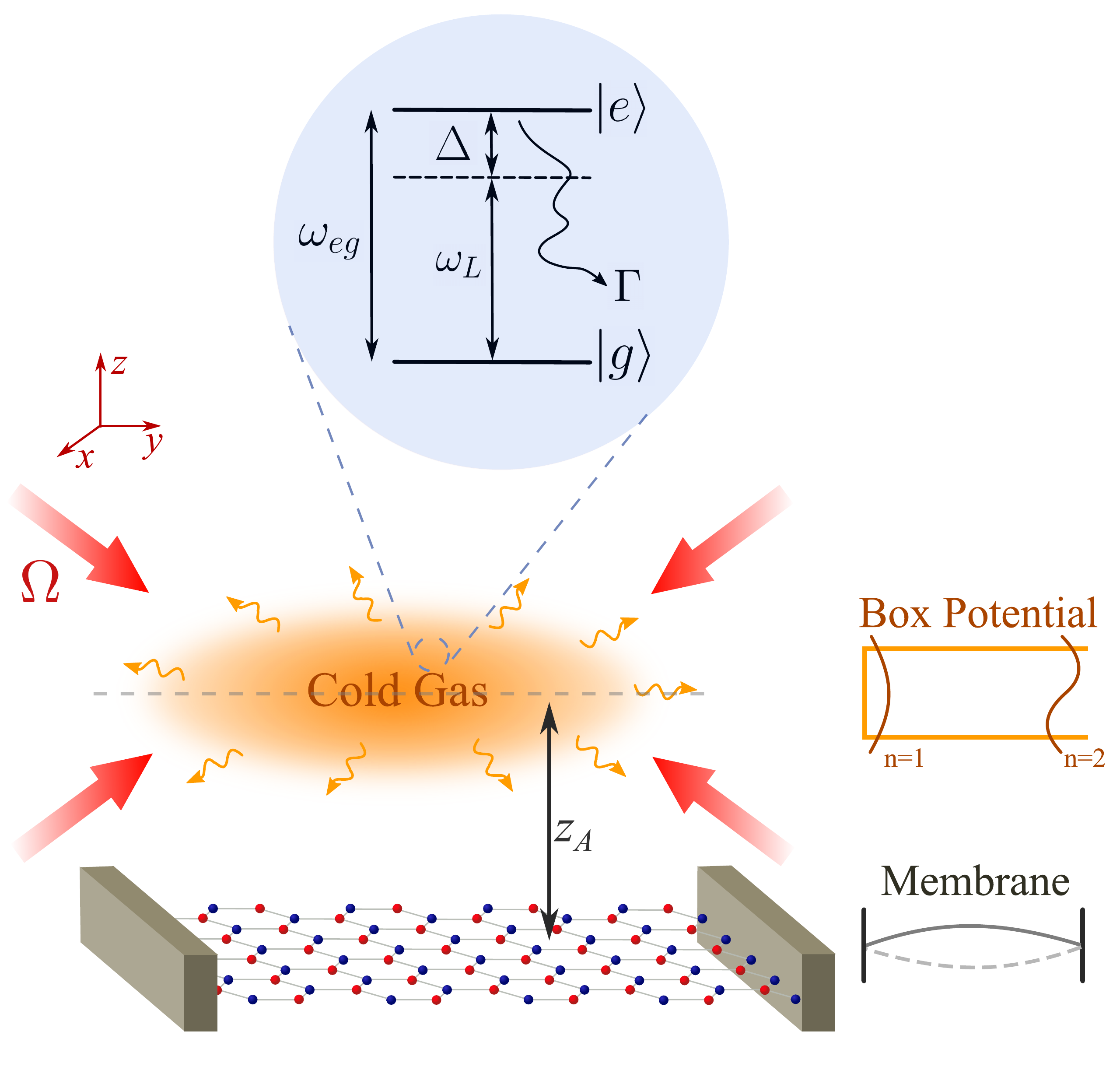}
  \caption{(Colour online) Scheme of the experimental setup of a quantum gas at a distance $z_A$ from a graphene sheet. The motion of the particles is coupled to the flexural modes of graphene via Casimir-Polder interactions. The cooling of the vibrational modes of the gas can be done with the help of the cooling laser with Rabi frequency $\Omega$. For our calculations, we have chosen an atomic cloud of rubidium ($^{87}$Rb).\label{Fig:ExperimentalSetup}}
 \end{figure}

There has been growing interest in exploring carbon nanotubes held at positive voltage to capture an ionize individual cold atoms \cite{PRA79_043403_2009, PRL104_133002_2010}, but also to explore the dispersion interactions between Bose-Einstein condensates and carbon nanotubes \cite{NewJPhys15_073009_2013, NatNano7_515_2012}, and laser-controlled ultracold (Rydberg) atoms and graphene setups that may be used to create ripples on demand \cite{PRA88_052521_2013}. The dispersion
\begin{widetext}
\begin{figure*}[t!]
\centering
  \includegraphics[width=18cm]{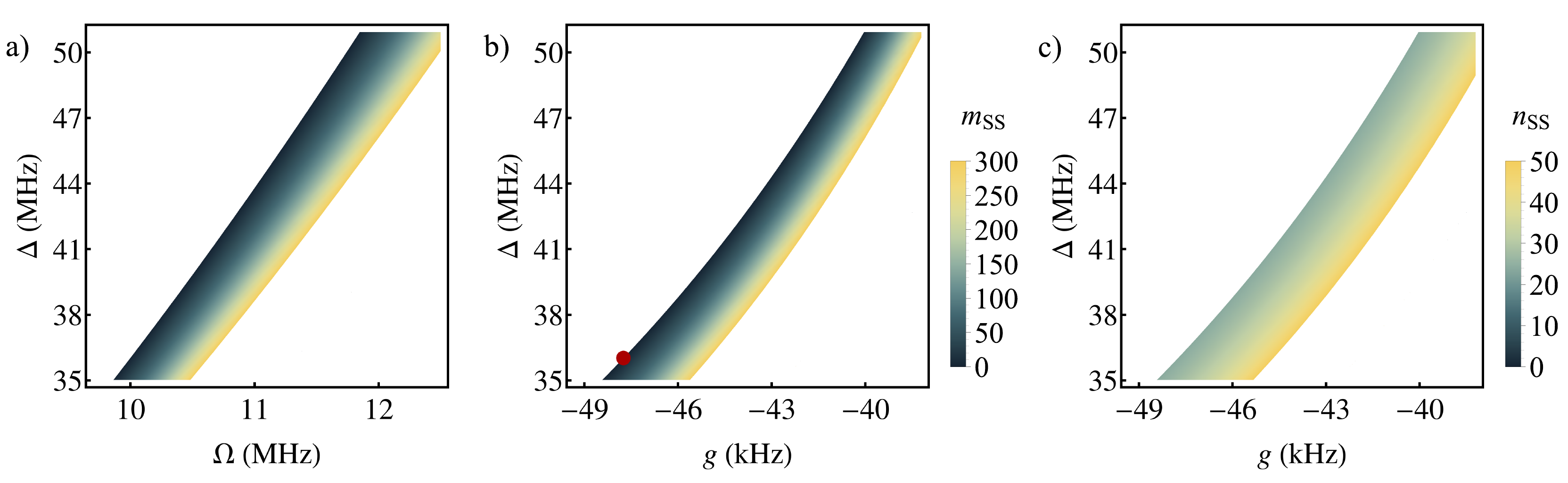}
  \onecolumngrid \caption{(Colour online) Density plot of the stationary state flexural mode number $m_\text{SS}$ a) and b) and $n_\text{SS}$ c) for the parameters $\eta = 0.25$, $\Gamma = 6.07$~MHz, $\nu = 2.7$~MHz, $\omega_\text{ph} = 477$~Hz, where in a) and c) $\Omega = 10$~MHz and for b) $g= -47.7$~kHz. The red point in pannel b) marks the coordinate $g=-47.7$~kHz and $\Delta = 36$~MHz, where $m_\text{SS} \sim 0.15$ for which the time evolution in Fig.~\ref{Fig:decayplot} is calculated, and corresponds to $n_\text{SS} \approx \pc{\exp \pr{\hbar \omega_\text{ph} /\pc{ k_B T_{\text{atoms}}}} -1}^{-1} \sim 24$ for which the temperature is $T_{\text{atoms}} =560$~nK. \label{Fig:mSSplusnSSplot}}
\end{figure*}
\end{widetext}
 forces arising from  quantum fluctuations of the electromagnetic field between cold atoms and a carbon nanotube, commonly known as Casimir-Polder (CP) interaction \cite{acta2008}, has been experimentally measured and pointed as potentially useful for applications in quantum sensing and quantum information \cite{schneeweiss}. Immersing nanotubes in cold atom clouds has also been suggested as a passive sympathetic cooling method \cite{weiss}. In a recent study, the authors have theoretically examined the possibility of strong-coupling matter waves with a graphene sheet, paving the stage for non-destructive cold atom-graphene interfaces \cite{tercas_sofia}. 

In this Letter, we propose a novel scheme to approach the quantum limit of the zero-point flexural motion of a graphene sheet by sympathetic laser-cooling via vacuum forces. Our setup consists in a laser-cooled two-dimensional cloud of cold atoms that is placed near the membrane (a few $\mu$m). At this distance, the atoms and the flexural modes couple as a consequence the CP interaction, allowing the exchange of excitations without destroying the atomic cloud \cite{tercas_sofia}. By cooling the atomic motion with a far-detuned laser, we show that it is possible to effectively tailor the dissipation of the graphene membrane via vacuum fluctuations. Contrary to other optomechanical cooling protocols \cite{PRL113_027404_2014}, our method offers the possibility of discarding dilution refrigerators and allows to cool graphene to the quantum limit starting from room temperatures. Also, because it is non-destructive, it offers advantages with respect to Ref.~\cite{weiss} as it allows the {\it active} cooling of graphene zero-point motion to the quantum limit in steady-state. 

Our setup consists of a cold atomic gas confined in a two-dimensional box potential - in such a way that its transverse center-of-mass (phonon) modes are quantized (see Fig.~\ref{Fig:ExperimentalSetup}) - placed near a suspended graphene sheet (for our numerical calculations, we have chosen an atomic cloud of $^{87}$Rb \cite{RubidiumData}). The phonons then couple to the flexural (out-of-plane) modes of a graphene sheet via vacuum CP forces and the cooling laser drives D2 $^{87}$Rb transition. The aim of the laser cooling is to dissipate the atomic motion and, as a consequence, sympathetically cool down the flexural modes of the graphene sheet. In this manuscript, we will restrict our discussion to a single phonon mode coupled to single flexural mode.

The Hamiltonian of the system in Fig.~\ref{Fig:ExperimentalSetup} can be written as
\begin{eqnarray}
\hat{H} = \hat{H}_\text{at} + \hat{H}_\text{ph} +  \hat{H}_\text{flex} + \hat{H}_\text{L} + \hat{H}_\text{at-graph}.
\label{eq:Htotalsystem}
\end{eqnarray}
The first three terms are the energy of the electronic states of the atoms, their quantized vibrational modes and the fundamental mode of the graphene sheet. Here, the cloud is composed of atoms with ground state $\ket{g}$, excited state $\ket{e}$ and transition frequency $\omega_{eg}=\omega_e-\omega_g$ (see Fig.~\ref{Fig:ExperimentalSetup}); $\omega_\text{ph}$ and $\nu$ are the center-of-mass (phonon) excitation in the atomic cloud and the flexural mode energies, respectively. Then, we can explicitly write
\begin{eqnarray}
\hat{H}_\text{at} &=& \hbar \pc{\omega_e \ket{e}\bra{e} + \omega_g \ket{g}\bra{g}}, \\
\hat{H}_\text{ph} &=& \hbar \omega_\text{ph} \hat{a}^\dagger \hat{a}, \\
\hat{H}_\text{flex} &=& \hbar \nu \hat{f}^\dagger \hat{f},
\end{eqnarray}
where $\hat{a}$ and $\hat{f}$ are the phonon and the flexural bosonic annihilation operators.
 
The fourth term in Eq.~\eqref{eq:Htotalsystem} describes the coupling between the laser and the center-of-mass motion of the atoms, which can be written in the rotating wave approximation (RWA) as
\begin{eqnarray}
\hat{H}_\text{L} = \frac{\hbar}{2}\Omega \pc{\sigma_- \hat{D} \pc{i \eta} e^{i \omega_L t}+ \text{H.c.}},
\end{eqnarray}
where $\omega_L$ is the laser frequency, $\Omega = \mathbf{d}_{ge} \cdot \boldsymbol{\epsilon} E_0 / \hbar$ is the Rabi frequency of the transition $\ket{g} \to \ket{e}$ (with $E_0$ being the electric field amplitude, $\mathbf{d}_{ge}$ the dipole operator and $\boldsymbol{\epsilon}$ the laser polarization). We have introduced the lowering and raising operators $\sigma_- = \pc{\sigma_+}^\dagger = \ket{g}\bra{e}$ and the displacement operator $\hat{D} (i \eta) \equiv e^{-i \eta \pc{\hat{a}^\dagger + \hat{a}}}$, where $\eta=\omega_{\rm rec}/\omega_{\rm ph}$ is the Lamb-Dicke parameter and $\omega_{\rm rec}$ is the recoil frequency. Taking advantage that the experimental parameters $\Omega, \; \omega_\text{ph}$ and $\Delta = \omega_{eg}- \omega_L$ are much smaller than the optical frequency, we can perform the atom-laser RWA Hamiltonian as
\begin{eqnarray}
\hat{H}_\text{at} + \hat{H}_\text{L} =\hbar \Delta \sigma_+ \sigma_-  +
\frac{\hbar}{2}\Omega \pc{\sigma_- \hat{D} \pc{i \eta}+ \text{H.c.}}.
\end{eqnarray} 
Finally, the last term in Eq.~\eqref{eq:Htotalsystem} is the Hamiltonian describing the interaction between the atoms and the graphene sheet that can be obtained by expanding the CP potential around the equilibrium position at first order in the displacement operator and averaging over the atomic density (see \cite{SuppMat} for details), which reads
\begin{align}
\hat{H}_\text{at-graph} &= \hbar \pc{ \omega^{\ket{e}} \ket{e} \bra{e} + \omega^{\ket{g}} \ket{g} \bra{g}} \hat{a}^\dagger \hat{a} \hat{\mathcal{T}}.
\end{align}
Here, $\omega^{\ket{i}}$ is Fourier transform of the CP potential of the electronic state $\ket{i}$, and $\hat{\mathcal{T}} \equiv 1 + i 2 q_0 \sqrt{\hbar/ (2 m \nu)} \pc{\hat{f} +\hat{f}^\dagger}$ is the translation operator. We restrict the discussion to the fundamental mode $q_0\equiv 2 \pi/L$, with $L$ representing the size of the graphene sheet (here considered squared, for definiteness).

\begin{figure}[t]
\centering
  \includegraphics[width=8.5cm]{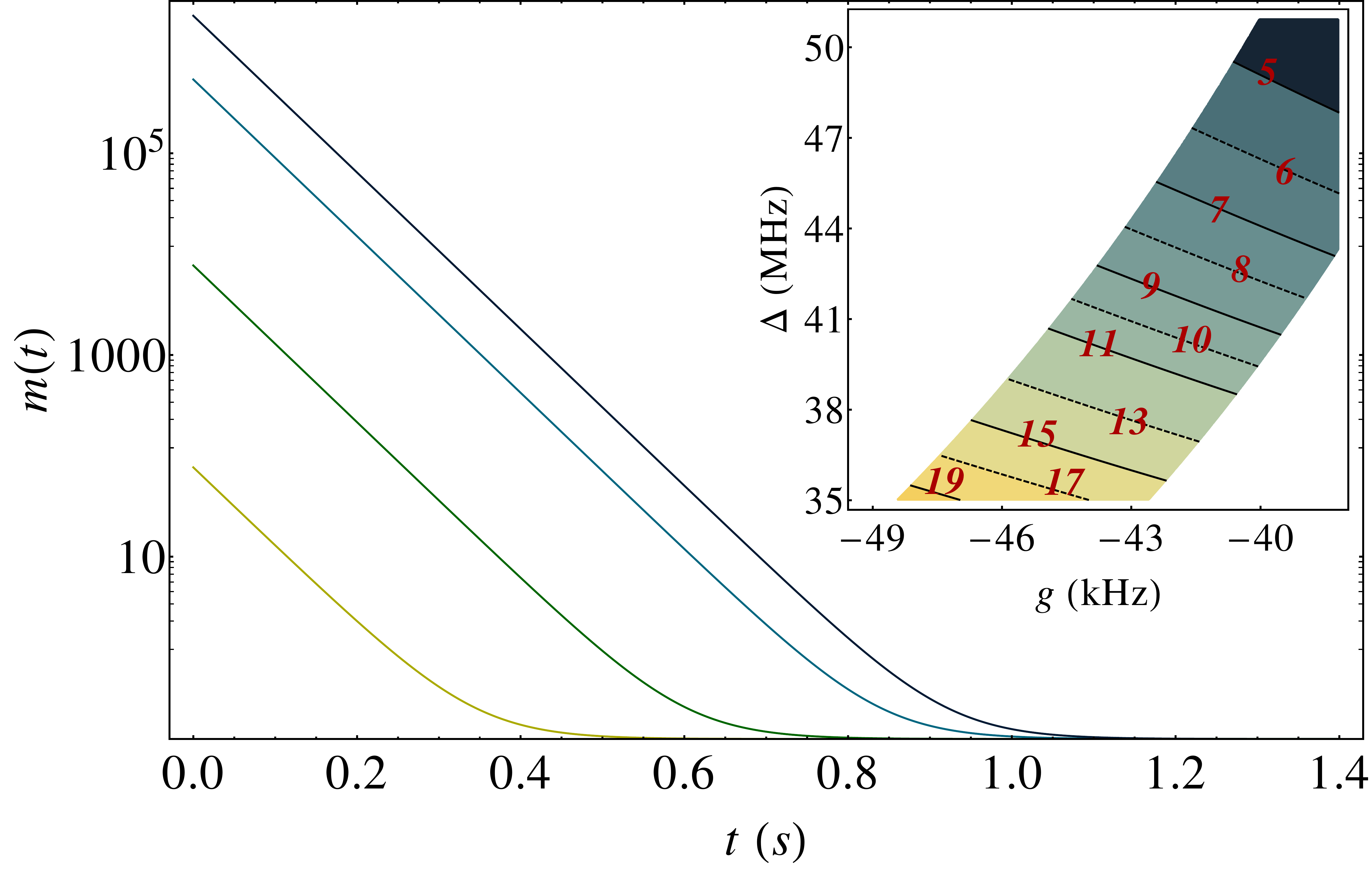}
  \caption{(Colour online) Logarithmic plot of the time evolution of the mean flexural mode number $m$ for the same parameter as Fig.~\ref{Fig:mSSplusnSSplot}b), but with $g=-47.7$~kHz and $\Delta=36$~MHz. The different lines represent different starting temperature, from bottom up $T= 0.01,\, 1,\, 70, \, 300$~K. For these parameter the $m_\text{SS} = 0.15$ which corresponds to a final temperature of $T_\text{graph} = 60~\mu$K. The inset shows a contour plot of $\gamma_\text{eff}$ as a function of $g$ and $\Delta$ for the same parameters as Fig.~\ref{Fig:mSSplusnSSplot}b). \label{Fig:decayplot}}
 \end{figure}

To minimize spontaneous emission from the excited electronic state of the atoms, we assume that the detuning $\Delta$ is much larger than all other system parameters, $\Delta \gg \Omega, \, \omega_\text{ph}, \nu, \, \omega^{\ket{i}}, \, \Gamma.$
Under such conditions, we are able to adiabatic eliminate the electronic states from the time evolution and obtain an effective master equation which reduces the evolution to the ground-state dynamics \cite{PRA85_032111_2012}. The dynamics of the reduced density matrix $\hat \rho$ is thus governed by the following master equation
\begin{equation}
\dot{\hat \rho} = -\frac{i}{\hbar} \pr{\hat{H}_{\rm eff}, \hat \rho} +\mathcal{L}_{\rm eff}(\hat \rho),
\label{eq_lind}
\end{equation}
where the effective Hamiltonian and Lindblad operators are given by (see more details about these calculations in \cite{SuppMat}),
\begin{align}
\hat{H}_\text{eff} &= \hbar  \omega \hat{a}^\dagger \hat{a} + \hbar \nu \hat{f}^\dagger \hat{f} + i \hbar g \, \hat{a}^\dagger \hat{a} \pc{\hat{f}^\dagger + \hat{f}} \nonumber \\
&\quad+ i \hbar \xi \pc{\hat{a}^\dagger + \hat{a}} , \label{eq:Heff1} \\
\mathcal{L}_\text{eff} \pc{\hat{O}} &= \frac{\gamma}{2} \px{2 \hat{a}^\dagger \hat{O} \hat{a} - \hat{O}\hat{a}^\dagger \hat{a} - \hat{a}^\dagger \hat{a} \hat{O}}.
\label{eq:Leff}
\end{align}
Here, we have defined the reduced quantities $\omega= \omega_\text{ph} - \eta^2 \hbar \Omega^2 \Delta/\pc{4 \Delta^2 + \Gamma^2} + \omega^{\ket{g}}$, $\xi = \eta \hbar \Omega^2 \Delta \pc{4 \Delta^2 + \Gamma^2}$, $g= 2 q_0 \sqrt{\hbar / (2 m \nu)} n_0 \omega^{\ket{g}} $ with $n_0$ being the atomic density, $\gamma =\Gamma  \eta^2 \Omega^2 / \pc{\Gamma +4 \Delta ^2}$ with $\Gamma$ denoting the atomic spontaneous emission rate. Moreover, $\hbar \omega^{\ket{g}} = \pi C_4 q_0 n_0 K_1 (q_0 z_A) /z_A$ is the Fourier transform of the CP potential $U^{\ket{g}}_{\text{CP}}$, with $K_1 (x)$ denoting the modified Bessel function of the second kind \cite{tercas_sofia}. In the present work, we have considered the retarded limit, that corresponds to the situation where the atom-surface distance $z_A$ is large when compared to the effective transition wavelength,  $z_A \gg c / \omega_\text{eg} $ \cite{acta2008}. In this situation, $U^{\ket{g}}_{\text{CP}} = C_4 / z_A^4$, where one finds $C_4 = −14.26$~Hz$\mu$m$^4$ for graphene interfacing with $^{87}$Rb in the ground state -- in the numerical calculations, the coupling chosen to produce the results in Fig.~\ref{Fig:decayplot} corresponds to $z_A \approx 0.1~\mu$m.

Analytical solutions to Eq. (\ref{eq_lind}) are possible by linearizing the phonon operator around its average amplitude $\av{a}\equiv\alpha$ \cite{SCPMA58_050302_2015}. The displaced phonon operator can then be defined as $ \hat{a} \to  \alpha+\delta \hat{a}$, where $\av{\delta\hat{a}} =0$ and the total phonon number given by $\hat n=\md{\alpha}^2 + \av{\delta \hat{a}^\dagger \delta \hat{a}}$ \cite{PRL108_153601_2012}. 
Inserting these expressions into the Eqs.~\eqref{eq:Heff1} and \eqref{eq:Leff}, we find linearized forms of both the Hamiltonian and Lindblad operators \cite{SuppMat}
\begin{align}
\tilde{H} &= \hbar  \omega \delta \hat{a}^\dagger \delta \hat{a}+ \hbar \omega \md{\alpha}^2 +\hbar \nu \hat{f}^\dagger \hat{f} + i \hbar g \,\md{\alpha}^2 \pc{\hat{f}^\dagger + \hat{f}} \nonumber \\
&\quad + i \hbar g \, \pc{\delta\hat{a}^\dagger \alpha + \delta\hat{a} \alpha^*} \pc{\hat{f}^\dagger + \hat{f}} - i \hbar \xi \pc{\alpha^* - \alpha},\label{eq:totH}\\
\tilde{\mathcal{L}} \pc{\hat{O}} &= \frac{\gamma}{2}  \px{2 \delta\hat{a}^\dagger \hat{O} \delta\hat{a} - \hat{O}\delta\hat{a}^\dagger \delta\hat{a} - \delta\hat{a}^\dagger\delta \hat{a} \hat{O}} \nonumber \\
&\quad + \frac{\gamma}{2} \pc{ \alpha\pr{\delta\hat{a}^\dagger, \hat{O}}+ \alpha^* \pr{\hat{O}, \delta\hat{a}}}.
\end{align}
Having determined the linearized dynamics, we solve the master equation to obtain the average number of phonons $n=\av{\hat n}$ and flexurons $m = \av{\hat{f}^\dagger \hat{f}}$, and the corresponding effective cooling rate $\gamma_{\text{eff}}$. Making use of the property $\av{\dot{O}} = \Tr \pr{\hat{O} \dot{\rho}}$, Eq. (\ref{eq_lind}) simply yields
\begin{align}
\av{\dot{O}} &= -\frac{i}{\hbar} \av{\pr{\hat{O}, \hat{\tilde{H}}}}+ \frac{1}{2} \gamma
\av{2 \hat{a}^\dagger \hat{O} \hat{a} - \hat{O} \hat{a}^\dagger  \hat{a} - \hat{a}^\dagger \hat{a} \hat{O}  } \nonumber \\
&\quad + \frac{1}{2} \gamma \av{ \alpha\pr{\hat{a}^\dagger, \hat{O}}+ \alpha^* \pr{\hat{O}, \hat{a}}}
. \label{eq:master}
\end{align}
Finally, from Eqs.~\eqref{eq_lind} and \eqref{eq:master}, we obtain the time evolution for the mean flexural number $m(t)$, the mean phonon number $n(t) $, and for the coherences $\hat{k}_1 = i \pc{\hat{f}^\dagger + \hat{f}}$, $\hat{k}_2 = \hat{f}^\dagger - \hat{f}$, $
\hat{k}_3 = \alpha \hat{a}^\dagger + \alpha^* \hat{a}$, $\hat{k}_4 = i \pc{\alpha \hat{a}^\dagger - \alpha^* \hat{a}}$, $\hat{k}_5 = i \pc{\alpha \hat{a}^\dagger + \alpha^* \hat{a}} \pc{\hat{f}^\dagger + \hat{f}}$, $\hat{k}_6 =  \pc{\alpha \hat{a}^\dagger + \alpha^* \hat{a}} \pc{\hat{f}^\dagger - \hat{f}} $, $\hat{k}_7 =  \pc{\alpha \hat{a}^\dagger - \alpha^* \hat{a}} \pc{\hat{f}^\dagger + \hat{f}}$, $\hat{k}_8 = i \pc{\alpha \hat{a}^\dagger - \alpha^* \hat{a}} \pc{\hat{f}^\dagger - \hat{f}}$, $\hat{k}_9 = \hat{f}^{\dagger2} + \hat{f}^2$, $ \hat{k}_{10} = i \pc{ \hat{f}^{\dagger 2} - \hat{f}^2}$, $\hat{k}_{11} = \pc{\alpha \hat{a}^{\dagger}}^{2} + \pc{\alpha^* \hat{a}}^2 $ and $\hat{k}_{12} = i \pc{ \pc{\alpha \hat{a}^{\dagger}}^{2} - \pc{\alpha^* \hat{a}}}^2$ .

In order to calculate the stationary flexural number $m_{\rm SS}$, we calculate Eq.~\eqref{eq:master} for all the coherences, $\av{\dot{k}_i}$ and occupation numbers, $\av{\dot{m}}$ and $\av{\dot{n}}$, and set them equal to zero. This gives a set of 14 equations which can be easily solved to find $m_{\rm SS}$, $n_{\rm SS}$ and $k^{\rm SS}_{i}$ (see details of these calculations in \cite{SuppMat}).
We are especially interested in seeing how the value of the stationary state $m_\text{SS}$ evolves with the tuneable parameters of the experiment $g$, that we can tune by changing the atom-surface distance, $\Delta$ and $\Omega$ that we can tune in the cooling process. These results are shown in Fig.~\ref{Fig:mSSplusnSSplot}a) and \ref{Fig:mSSplusnSSplot}b). With an appropriate choice of $\Delta$, $\Omega$ and $g$, it is shown to be possible to have steady states with number of flexural modes lower to 10, which corresponds to temperatures lower than miliKelvin. 
In the set of parameters chosen $\eta = 0.25$, $\Gamma = 6.07$~MHz, $\nu =2.7$~MHz, $\Omega = 10$~MHz, $\omega_\text{ph} = 477$~Hz, $g=-47.7$~kHz and $\Delta =36$~MHz (the red point marked in Fig.~\ref{Fig:mSSplusnSSplot}b)), $m_\text{SS} \sim 0.15$ that corresponds to a temperature of $T_{\text{graph}} \simeq 60 ~\mu$K. This lies at least two orders of magnitude below the experimental record obtained with dilution refrigerators \cite{PRL113_027404_2014}. Correspondingly, the atomic motion is limited to $n_{\rm SS}\sim 24$ (see Fig.~\ref{Fig:mSSplusnSSplot}c)) for which we find $T_{\text{atoms}} =560$~nK. This is consistent with the Lamb-Dicke approximation for a two-dimensional gas confined in a box potential of frequency $\omega_z$ of the order of a few kHz \cite{PRL110_200406_2013, NatureComm6_6162_2015}.

For an analytical estimate of the effective cooling rate $\gamma_\text{eff}$, we take that the mean flexural mode number $m$ evolves on a relatively slow time scale comparative to all other variables, which dynamics can be adiabatically eliminated by setting $\av{\dot{k}_i} = 0$ and $\av{\dot{n}}=0$. As such, we find a cooling dynamics of the type $m (t) = a e^{-\gamma_\text{eff} t} + m_\text{SS} $. In the parameter regions for which the cooling is stable (i.e. for which the rate $\gamma_\text{eff}$ is positive, corresponding to the shadowed regions in Fig. \ref{Fig:mSSplusnSSplot}), cooling is possible irrespective of the initial graphene temperature. Fig.~\ref{Fig:decayplot} compares the sympathetic graphene laser-cooling dynamics for different initial occupation numbers $m(0)=\pc{\exp \pr{\hbar \nu / \pc{k_B T_{\text{graph}}} }-1}^{-1}$. The exponential reduction of the occupation flexural number only slows downs until $m$ reaches its stationary state value and $\gamma_\text{eff} \approx 2.87$~Hz. Even starting from room temperature, the stationary state is reached within a experimentally reasonable time scale ($\tau_{\rm eff}\simeq 0.35$ s). 
Please note that although the CP potential is temperature dependent, for the results in Fig.~\ref{Fig:decayplot} we have considered a fixed interaction $g=-47.7$~kHz for any initial temperature, this coupling could be experimentally achieved by tuning the atom-surface distance. Also, at higher temperatures, it can happen that we would need to include the effect of secondary mechanical modes. The cooling of the fundamental mode would depend upon the difference between the two or more mechanical resonance frequencies, one expects that only when this difference is large than the effective damping of the secondary mode, the cooling process not to be affected \cite{NewJPhys10_095009_2008}.

Another important remark about the nature of the coupling is in order. In principle, two important physical effects could constitute a limitation to the present cooling protocol. First, the CP potential depends on the temperature of the two media (graphene sheet and atoms) \cite{PRL100_253201_2008,PRA80_022901_2009, PRL104_223003_2010}. However, this dependence is negligible for the ultracold atomic temperatures. Second, blackbody radiation effects could also lead to a spurious heating of the atoms \cite{PRA56_5100_1997}. However, we have shown that this effect is not relevant for distances of a few $\mu$m \cite{tercas_sofia}.
 
In conclusion, in this manuscript, we have shown that is possible via the CP interactions between an atomic gas and a graphene sheet to cool the mechanical out-of-plane vibrations, by laser cooling the phonon excitations in the quantum gas. Sympathetic laser cooling via the vacuum forces is shown to reach temperatures of $\sim 60~\mu$K, about 100 times cooler than the temperature reached by the usual optomechanical methods ($\sim$~60~mK).
Cooling a membrane to its ground state creates a path towards new hybrid systems where the motion of mechanical resonators can be coupled with other quantum systems. One can imagine in these systems that the membrane could act as transducers providing coupling between, for instance, photons or spins. Such transducers could play an important role in quantum networks. For the condensate matter physics point of view, the interest lies in cooling the mechanical modes of graphene and, as so, controlling the transport properties. Since the mobility of carriers in graphene membranes are extremely sensitive to temperature, flexural phonons are the leading scattering mechanism for temperatures higher than few Kelvin, by cooling the mechanical modes of a graphene membrane it would then be possible to decrease the resistivity of the membrane.

The authors would like to thank the support from Funda\c{c}\~{a}o para a Ci\^{e}ncia e a Tecnologia (Portugal), namely SR through the projects UID/EEA/50008/2013, IT/QuSim, and HT through the scholarship SFRH/BPD/110059/2015. SR and HT would like to acknowledge fruitful discussions with J.D.~Rodrigues.

\newpage

\begin{center}
{\bf \large Supplementary Material}
\end{center}

\setcounter{equation}{0}
\section{Casimir-Polder Physics}

For planar structures, the Casimir-Polder (CP) potential of an atom in the ground state at a
distance $z_A$ away from the macroscopic body with permittivity $\varepsilon(\omega)$ can be written as \cite{acta2008_1}
\begin{align}
U_{\mathrm{CP}}^{\ket{g}} \pc{z_{A}} &= \frac{\hbar \mu_{0}}{8 \pi^{2}} \int_{0}^{\infty} d \xi \xi^{2} \alpha_{at} \pc{i \xi} \int\limits_{0}^{\infty} d k_{\parallel}
 \frac{e^{-2 k_{\parallel} \gamma_{0z} z_{A}} }{\gamma_{0z}} \nonumber \\
 &\quad \times \left[
\mathrm{R}_{\mathrm{TE}} + \mathrm{R}_{\mathrm{TM}} \left( 1- \frac{2 k_{\parallel}^{2} \gamma_{0z}^{2} c^{2} }{\xi^{2}} \right) \right]    
\label{eq:Ucp_1}
\end{align}
and for an atom in an energy eigenstate $|n\rangle$ as
\begin{gather}
U_{\mathrm{CP}}^{\ket{n}} \pc{z_{A}} = \frac{\hbar \mu_{0}}{8 \pi^{2}} \int_{0}^{\infty}
d \xi\, \xi^{2} \alpha_n(i \xi) \int\limits_{0}^{\infty} d k_{\parallel}
\frac{e^{-2 k_{\parallel} \gamma_{0z} z_{A}} }{\gamma_{0z}} \nonumber \\
\times \left[ \mathrm{R}_{\mathrm{TE}} + \mathrm{R}_{\mathrm{TM}} \left( 1-
\frac{2 k_{\parallel}^{2} \gamma_{0z}^{2} c^{2} }{\xi^{2}} \right) \right] 
\nonumber \\
+ \frac{\mu_{0}}{4 \pi} \sum_{k \neq n} \omega_{n k}^{2} \mathbf{d}_{n k}
\cdot \mathbf{d}_{k n} \int_{0}^{\infty} d \kappa_{0z} e^{-2 \kappa_{0z} z_A}
\nonumber \\
 \times \mathrm{Re} \pr{\mathrm{R}_{\mathrm{TE}} +
\mathrm{R}_{\mathrm{TM}} \pc{1 + \frac{2 \kappa_{0z}^{2}
c^{2}}{\omega^{2}}}  }.  
\label{eq:Ucp}
\end{gather}
where $\gamma_{iz} = \sqrt{1+\varepsilon_i (i\xi)\xi^{2}/(c^{2} k_\|^2)}$, $\kappa_{0z} = \sqrt{ k_\|^2+\omega^2/c^2}$, $\omega_{ij}$ $\pc{\mathbf{d}_{ij}}$ is the transition frequency (dipole moment) and $\alpha_n(\omega)$ is the atomic polarizability defined by
\begin{equation}
\mathbf{\alpha}_n(\omega) = \lim_{\varepsilon \rightarrow 0} \frac{2}{\hbar}
\sum_{k \neq n} \frac{\omega_{kn}  \mathbf{d}_{n k}
\cdot \mathbf{d}_{k n}}{\omega_{kn}^2-\omega^2-i\omega\varepsilon}\,.
\label{eq:atomicpol}
\end{equation}
The first term in Eq.~(\ref{eq:Ucp}) describes the nonresonant part of the CP potential, recognisable by the integration along the imaginary frequency axis, $\omega=i\xi$, whereas the second term is related to resonant  photon exchange between the atom and the graphene sheet. Eqs~\eqref{eq:Ucp_1}, \eqref{eq:Ucp} are strictly valid only at zero temperature.  We are assuming potential experimental setups that could be performed at sufficiently low temperatures for thermal excitations to only play a subordinant role; in addition, the distance of those atoms from the graphene sheet will be much smaller than the thermal wavelength $\lambda_T=hc/(k_B T)$. In situations in which either assumption fails to hold, a replacement of the frequency integral by a Matsubara sum,
\begin{equation}
\frac{\hbar}{\pi}\int\limits_0^\infty d\xi\,f(i\xi)\mapsto
2k_BT\sum\limits_{n=0}^\infty \left(1-\frac{1}{2}\delta_{0n} \right)f(i\xi_n)
\,,
\end{equation}
with Matsubara frequencies $\xi_n=2\pi k_BTn/\hbar$ \cite{thermalCP}, has to be employed.
Thermal corrections become important only for $k_B T \gtrsim \Delta$, where $\Delta$ is the gap parameter of quasiparticle excitations \cite{PRA86_012515_2012}. 
At finite temperature, the potential is well approximated by inserting the temperature-dependent reflection coefficients in  the lowest term in the Matsubara sum ($j=0$) while keeping the zero-temperature coefficients for all higher Matsubara terms \cite{PRB82_155459_2010, PRB85_195427_2012}.

Due to graphene's unique electronic structure, a full calculation of its electromagnetic reflection coefficients is in fact possible from first principles. Using the Dirac model for the description of the dynamics of quasiparticles in graphene at zero temperature in external electromagnetic fields, one can, from the boundary conditions of the fields, find the reflection coefficients for given values of the mass gap $m$ and chemical potential $\mu$. For simplicity we will set $m =\mu=0$ (perfect Dirac cone) for which the difference between this approximation for suspended graphene samples ($m, \mu \sim 0.01$~eV) is less than $1\%$ \cite{PRB84_035446_2011}. One then arrives at the reflection coefficients of a free standing graphene sheet in vacuum as
\begin{align}
 \mathrm{R}_{\mathrm{TM}} &= \frac{4 \pi \alpha \sqrt{k_{0}^{2} + k_{\parallel}^{2}} }{4 \pi \alpha \sqrt{k_{0}^{2} + k_{\parallel}^{2}} + 8 \sqrt{k_{0}^{2} + \tilde{v}^{2} k_{\parallel}^{2}}}, \\
 \mathrm{R}_{\mathrm{TE}} &= - \frac{4 \pi \alpha\sqrt{k_{0}^{2} + \tilde{v}^{2} k_{\parallel}^{2}} }{4 \pi \alpha \sqrt{k_{0}^{2} + \tilde{v}^{2} k_{\parallel}^{2}} + 8 \sqrt{k_{0}^{2} + k_{\parallel}^{2}}}
\label{eq:reflcoefG}
\end{align}
where we define $k_{0}^{2} = \xi^{2}/c^{2}$ and $\tilde{v}=(300)^{-1}$, and $\alpha=1/137$ is the fine structure constant. 

\subsection{Interaction Hamiltonian of the atomic gas-graphene system}

In general, the CP potential $U_{\text{CP}} \pc{\mathbf{R}_A}$ , where $\mathbf{R}_A = \pc{x_A, y_A, z_A}$ has a part that depends only on $z$, that for planar surfaces gives the usual CP force $U_{\text{CP}} (z_A)= \frac{C_n}{z_A^n}$ and one that depends both on $x_A$ and $y_A$, that gives a lateral component to the CP force due to the lack of translational invariance, (where we will assume $\mathbf{r}_A = (x,y)$). The later as already been calculated in Refs.~\cite{PRL100_040405_2008, JPhysA41_164028_2008}. The potential for a very smooth surface can be obtained from the planar case by merely taking the local atom-surface distance
\begin{align}
U (x,y,z) &\simeq U^0_{\text{CP}} \pc{z_A} - u \pc{\mathbf{r}_A} \frac{d U^0_{\text{CP}} \pc{z_A}}{d z_A} \nonumber\\
&= \frac{C_n}{z_A^n} + u \pc{\mathbf{r}_A} n \frac{C_n}{z_A^{n+1}}.
\end{align}

To obtain the interaction energy between atoms in the atomic gas and the corrugated membrane, one needs to sum the CP potential weighted by the atomic density over all space and over the area of the membrane, that is,
\begin{eqnarray}
\hat{H}_\text{int} &=& \int \frac{d \mathbf{R}}{\mathcal{V}} \int d \mathbf{R}' \; \hat{n} \pc{\mathbf{R'}} U \pc{\mathbf{R'-\mathbf{R}}} 
\end{eqnarray}
Since this is a convolution integral, it is more convenient to transform it into the Fourier domain where it becomes a separable sum of products of atomic and graphene variables. This yields,
\begin{eqnarray}
\hat{H}_\text{int} &= \int \frac{d \mathbf{R}}{\mathcal{V}} \int d \mathbf{R}' \; \hat{n} \pc{\mathbf{R'}} \px{U^0_{\text{CP}}  - u \pc{\mathbf{r}' - \mathbf{r}} \nabla U^0_{\text{CP}}} \nonumber \\
&\to\quad 
 \sum_q \hat{a}^\dagger \hat{a} \, U^0_q + i \sum_q \hat{a}^\dagger \hat{a} \, \mathbf{q} \cdot \hat{\mathbf{u}}_\mathbf{q} U^0_q,
\end{eqnarray}
where $U_q = \pi C_4 q n_0 K_1 (q d) /d$ is the Fourier transform of the potential $U^{\ket{g}}_{\text{CP}}$, with $n_0$ being the atomic density and $K_1 (x)$ gives the modified Bessel function of the second kind \cite{HugoMe2015}.

We can then express the interaction energy as
\begin{eqnarray}
\hat{H}_\text{int} \equiv \hbar \sum_\mathbf{q} w_\mathbf{q} \hat{a}^\dagger \hat{a} \hat{\mathcal{T}}_\mathbf{q},
\end{eqnarray}
where we define $w_\mathbf{q} \equiv U_\mathbf{q} / \hbar \sqrt{\mathcal{V}}$ and the translation operator of the graphene
\begin{eqnarray}
\hat{\mathcal{T}}_\mathbf{q} = 1 + i \mathbf{q} \cdot \hat{\mathbf{u}}_\mathbf{q}.
\end{eqnarray}
Expressing the phonon operator in the form
\begin{eqnarray}
\hat{\mathbf{u}}_\mathbf{q} = \frac{1}{\sqrt{2}} \sum_{\sigma} \phi_{\mathbf{q},\sigma} \pc{\mathbf{r}} \mathbf{e}_\sigma \pc{\hat{f}_{ \sigma} + \hat{f}^\dagger_{\sigma}}
\end{eqnarray}
with two polarizations $\sigma=(x,y)$ and satisfying the normalization condition $\av{\phi_\mathbf{q}, \phi_{\mathbf{q}'}} = \hbar / \pc{M \nu} \delta_{\mathbf{qq}'}$ where $M$ is the membrane mass.

In our setup, we will perform laser cooling meaning that we need to account for both electronic states $\ket{g}$ an $\ket{e}$ of the trapped particle, so we transform
\begin{align}
\hat{a}^\dagger &\to \mathcal{A}^{\ket{e}*} \hat{a}^\dagger\ket{e} \bra{e} +
 \mathcal{A}^{\ket{g}*} \hat{a}^\dagger \ket{g} \bra{g}, \\
 \hat{a}^\dagger &\to \mathcal{A}^{\ket{e}} \hat{a} \ket{e} \bra{e} +
 \mathcal{A}^{\ket{g}} \hat{a} \ket{g} \bra{g}, \\
 \hat{a}^\dagger \hat{a} &\to  \hat{a}^\dagger \hat{a}\ket{e} \bra{e}
 + \hat{a}^\dagger \hat{a} \ket{g} \ket{g}.
\end{align}

In such a way that the interaction Hamiltonian becomes
\begin{align}
\hat{H}_\text{int} &= \hbar \sum_\mathbf{q} \pc{w^{\ket{e}}_\mathbf{q} \ket{e} \bra{e} + w^{\ket{g}}_\mathbf{q} \ket{g} \bra{g} } \hat{a}^\dagger \hat{a}  \hat{\mathcal{T}}_\mathbf{q} .
\end{align}

\subsection*{Graphene Properties}

It is well known that a free-floating graphene sheet would always crumple at room temperature, hence to perform our setup there is the need to support the graphene sheet on a trench. Measurements on layered graphene sheets of thickness between 2 and 8~nm have provided spring constants that scale as expected with the dimensions of the suspended section, and range from 1 to 5~N/m \cite{JVSTB25_6_2007}. Other experiments studied the fundamental resonant frequencies from electromechanical
resonators made of graphene sheets \cite{Science_315_490_2007}. For mechanical resonators under tension $T$ the fundamental resonance mode $f_{0}$ is given by 
\begin{eqnarray}
f_{0} = 2 \pi \px{\pr{A \sqrt{\frac{E}{\rho}} \frac{t}{L^{2}}}^{2} + A^{2} 0.57
\frac{T}{\rho L^{2} w t}}^{1/2}
\end{eqnarray}
where $E$ is Young's modulus, $\rho$ is the mass density; $t,\, w, \, L$ are the thickness, width and length of the suspended graphene sheet and $A$ is a clamping coefficient ($A$ is equal to 1.03 for doubly clamped beams and 0.162 for cantilevers). 
We assumed a doubly clamped sheet with a finite value for the tension $T=1$~nN. Tension between graphene and trenches is a random process depending on the production technique and the interaction with the substrate and for that reason very difficult to control \cite{Science_315_490_2007}. For these study we have used the known values for bulk graphite $\rho= 2200$~kg/m$^{3}$ and $E= 1.0$~TPa, for a graphene doubly clamped with $t=$0.3~nm, $L=$5~$\mu$m and $w=$5~$\mu$m. 

\section{Adiabatic elimination of the electronic states.}

To minimize spontaneous emission from the excited electronic state of the trapped particles, we assume that the detuning $\Delta$ is much larger than all other system parameters, $\Delta \gg \Omega, \, \omega_\text{phn}, \nu, \, \omega^{\ket{i}}, \, \Gamma.$
Considering that our system consist of two subspaces, one for the ground states and one for the decaying excited states. 
The dynamics of the system are Markovian such that the time evolution density operator $\rho$ can be described by a master equation of Lindblad form,
\begin{align}
\dot{\rho} = -\frac{i}{\hbar} \pr{\hat{H}, \rho} + \pc{\hat{L} \rho \hat{L}^\dagger - \frac{1}{2} \pc{\hat{L}^\dagger \hat{L}  \rho + \rho \hat{L}^\dagger \hat{L} } },
\end{align}
each $\hat{L}$ represents a source of decay, the spontaneous decay which to takes the system from excited to the ground electronic state. Following Ref.~\cite{PRA85_032111_2012_1}, by combining perturbation theory of the density operator and adiabatic elimination of the excited states we reduce the dynamics to an effective master equation involving only the ground state manifold with effective Hamiltonian and Lindblad operators
\begin{align}
\hat{H}_\text{eff} &= -\frac{1}{2} \hat{V}_- \pr{\hat{H}^{-1}_\text{NH} + \pc{\hat{H}^{-1}_\text{NH}}^\dagger} \hat{V}_+ + H_\text{g} \label{eq:getHeff}\\
\hat{L}_\text{eff} &= \hat{L} \hat{H}^{-1}_\text{NH} V_+
\label{eq:getLeff}
\end{align}
connecting only the ground state. $\hat{V}_+ \pc{\hat{V}_-}$ are the perturbative (de-)excitations of the system and $\hat{H}_g$ is the ground state Hamiltonian, $\hat{H}_\text{NH}$ is the non-Hermitian Hamiltonian of the quantum jump formalism
\begin{align}
\hat{H}_\text{NH} = \hat{H}_\text{e} - \frac{i}{2} \hat{L}^\dagger \hat{L},
\end{align}
with $\hat{H}_\text{e}$ being the Hamiltonian in the excited manifold.

The Hamiltonian of our system can be written as
\begin{align}
\hat{H} &= \hat{H}_\text{g} + \hat{H}_\text{e}, \\
\hat{H}_\text{g} &= \hbar \omega \hat{a}^\dagger \hat{a}
+ \hbar \nu \hat{f}^\dagger \hat{f} + \hbar  \omega^{\ket{g}} \ket{g} \bra{g} \hat{a}^\dagger \hat{a} \hat{\mathcal{T}},\\
\hat{H}_e &= \hbar \Delta \ket{e}\bra{e} +  \hbar \frac{\Omega}{2} \pr{\ket{g}\bra{e} \hat{D} \pc{i \eta} + \ket{e}\bra{g} \hat{D}^\dagger \pc{i \eta}} \nonumber \\
&\quad + \hbar \omega^{\ket{e}} \ket{e}\bra{e} \hat{a}^\dagger \hat{a} \hat{\mathcal{T}}.
\end{align}
We write the (de-)excitation as
\begin{align}
\hat{V}_+ = \hbar \frac{\Omega}{2} \ket{e} \bra{g} \hat{D}^\dagger \pc{i \eta}, \quad
\hat{V}_- = \hbar \frac{\Omega}{2} \ket{g} \bra{e} \hat{D} \pc{i \eta}.
\end{align}
To solve the terms involving the displacement operator, we suppose that our atoms are cooled enough to ensure that we are in the Lamb-Dicke regime. This means $\hat{D}^\dagger (i \eta) \simeq 1 + i \eta \pc{\hat{a}^\dagger + \hat{a}}$. Applying the rotating wave approximation, to extract the dynamics needed one gets
\begin{align*}
\hat{V}_+ = \hbar \frac{\Omega}{2} \ket{e} \bra{g} \pc{1 + i \eta \, \hat{a}}, \quad
\hat{V}_- = \hbar \frac{\Omega}{2} \ket{g} \bra{e} \pc{1 - i \eta \, \hat{a}^\dagger}.
\end{align*}
Spontaneous emission from the excited to the ground state at a rate $\Gamma$ is represented by the Lindblad operator, $\hat{L} = \sqrt{\Gamma} \ket{g} \bra{e}$.
Consequently, the non-Hermitian Hamiltonian is found to be
\begin{align}
\hat{H}_\text{NH} &= \hbar \pc{\Delta -i\frac{\Gamma}{2} + \omega^{\ket{e}} \hat{a}^\dagger \hat{a} \hat{\mathcal{T}}} \ket{e}\bra{e} \nonumber \\
&\quad+  \hbar \frac{\Omega}{2} \pr{\ket{g}\bra{e} \hat{D} \pc{i \eta} + \ket{e}\bra{g} \hat{D}^\dagger \pc{i \eta}}.
\label{eq:HNH}
\end{align}
By applying the effective Hamiltonian formula eq.~\eqref{eq:getHeff} and considering that parameter regime in which we are working we get to
\begin{align}
\hat{H}_\text{eff} &=  - \frac{\hbar \Omega^2 \Delta}{4 \Delta^2 + \Gamma^2} \ket{g} \bra{g} 
+ i \eta \frac{\hbar \Omega^2 \Delta}{4 \Delta^2 + \Gamma^2} \ket{g} \bra{g} \pc{\hat{a}^\dagger + \hat{a}} \nonumber \\
&\quad - \eta^2 \frac{\hbar \Omega^2 \Delta}{4 \Delta^2 + \Gamma^2} \ket{g} \bra{g} \hat{a}^\dagger \hat{a} +
\hbar  \omega_\text{phn} \hat{a}^\dagger \hat{a} + \hbar  \omega_{f} \hat{f}^\dagger \hat{f} \nonumber \\
&\quad+ \hbar \omega^{\ket{g}} \ket{g} \bra{g} \hat{a}^\dagger \hat{a} \hat{\mathcal{T}}. \label{eq:Heff}
\end{align}

By applying eq.~\eqref{eq:getLeff} together with $\hat{V_+}$, $\hat{H}_\text{NH}$ and $\hat{L}_\Gamma$ as specified above, 
we obtain a single effective Lindblad operator
\begin{align}
\hat{L}_\text{eff} &= \frac{\sqrt{\Gamma} \, \Omega}{2 \Delta -i \Gamma} \ket{g} \bra{g}  -i \frac{\sqrt{\Gamma} \, \Omega\, \eta }{2 \Delta -i \Gamma} \ket{g} \bra{g} \hat{a} .
\end{align}

\section{Linearisation procedure.}
Starting with a simplified view of our system we can write
\begin{align}
\hat{H}_\text{eff} &= \hbar  \omega \hat{a}^\dagger \hat{a} + \hbar \nu \hat{f}^\dagger \hat{f} + i \hbar g \, \hat{a}^\dagger \hat{a} \pc{\hat{f}^\dagger + \hat{f}} \nonumber \\
&\quad+ i \hbar \xi \pc{\hat{a}^\dagger + \hat{a}} ,  \\
\mathcal{L}_\text{eff} \pc{\hat{O}} &= \frac{\gamma}{2} \px{2 \hat{a}^\dagger \hat{O} \hat{a} - \hat{O}\hat{a}^\dagger \hat{a} - \hat{a}^\dagger \hat{a} \hat{O}}.
\end{align}
where we have defined $\omega= \omega_\text{phn} - \eta^2 \hbar \Omega^2 \Delta/\pc{4 \Delta^2 + \Gamma^2} + \omega^{\ket{g}}$, $\xi = \eta \hbar \Omega^2 \Delta \pc{4 \Delta^2 + \Gamma^2}$ and $g= 2 q_0 \sqrt{\hbar / (2 m \nu)} \omega^{\ket{g}} $.

Our Hamiltonian is clearly non-linear, which means that the Heisenberg-Langevin equations of motion will also be non-linear.  Linearisation of the Heisenberg-Langevin equations can be done in two ways  \cite{SCPMA58_050302_2015_1}, the Hamiltonian linearisation method, used often in quantum optomechanics and the equation linearisation method used in the semiclassical limit, we will perform the first.
The phonon operator will then be rewritten as a displacement transformation with an average amplitude $\alpha$ and a fluctuating part $\delta \hat{a}$ which represents the quantum fluctuations of the phonon mode around the average amplitude,
$ \hat{a} \to \delta \hat{a} + \alpha$, where now $\av{\hat{a}} =0$ and the total phonon number is given by $\md{\alpha}^2 + \av{\delta \hat{a}^\dagger \delta \hat{a}}$.

Following \cite{PRL108_153601_2012_1}, we start by considering the Hamiltonian without interactions $\hat{H} = \hbar  \omega \hat{a}^\dagger \hat{a} + \hbar \nu \hat{f}^\dagger \hat{f} + i \hbar \xi \pc{\hat{a}^\dagger + \hat{a}}$, the expectation value $\hat{a} (t)$ is given by
\begin{align}
\dot{\hat{a}} (t)&= - \frac{i}{\hbar} \pr{\hat{a}, \hat{H}_\text{eff}} = - i \omega \hat{a} + \xi,
\end{align}
such that $\alpha$ can be written as $\dot{\alpha} (t) = -i \omega \alpha + \xi$.

By applying a displacement operator $\hat{\mathcal{D}} = e^{ \alpha^* \delta \hat{a} - \alpha \delta \hat{a}^\dagger}$, to a state vector $\ket{\tilde{\psi}} = \hat{\mathcal{D}} \ket{\psi}$, we can derive the transformed Hamiltonian $\tilde{H}$ by expanding the Schr\"odinger equation $ i \hbar \partial_t \ket{\tilde{\psi}} = \pc{i \hbar \dot{\hat{\mathcal{D}}} + \hat{\mathcal{D}} \hat{H}_\text{eff} } \ket{\psi}$ which implies
\begin{align}
\tilde{H} &= i \hbar \pc{ \dot{\alpha}^* \delta \hat{a} - \dot{\alpha} \delta \hat{a}^\dagger} + \hat{\mathcal{D}}\hat{H}_{\text{eff}} \hat{\mathcal{D}}^\dagger,
\end{align}
having in mind the important property of the displacement operator $\hat{\mathcal{D}}G\pc{\hat{a}, \hat{a}^\dagger} \hat{\mathcal{D}}^\dagger = G \pc{\delta\hat{a} + \alpha, \delta\hat{a}^\dagger + \alpha^*}$, we find
\begin{align}
\tilde{H} &= \hbar  \omega \delta \hat{a}^\dagger \delta \hat{a}+ \hbar \omega \md{\alpha}^2 +\hbar \nu \hat{f}^\dagger \hat{f} - i \hbar \xi \pc{\alpha^* - \alpha}. \nonumber
\end{align}
Using the same notation one can now turn to the CP coupling and rewrite it as
\begin{align}
\hat{H}_\text{at-graph} &= i \hbar \delta \hat{a}^\dagger \delta \hat{a } \pc{\hat{f}^\dagger + \hat{f}} + i \hbar\md{\alpha}^2 \pc{\hat{f}^\dagger + \hat{f}} \nonumber \\
&\quad+ i \hbar g \pc{\alpha \delta \hat{a}^\dagger + \alpha^* \delta \hat{a}} \pc{\hat{f}^\dagger + \hat{f}} ,
\end{align}
the  first term corresponds to the three-wave mixing process and describes the intrinsic nonlinear process of our system, if we assume $\av{\delta \hat{a}} \ll \alpha$ one can ignore this term and obtain the linearised Hamiltonian. The last term is the average CP coupling where $g \md{\alpha}^2$ is the coupling strength depending on the average phonon number. So finally, we find our linearised Hamiltonian
\begin{align}
\tilde{H} &= \hbar  \omega \delta \hat{a}^\dagger \delta \hat{a}+ \hbar \omega \md{\alpha}^2 +\hbar \nu \hat{f}^\dagger \hat{f} + i \hbar g \,\md{\alpha}^2 \pc{\hat{f}^\dagger + \hat{f}} \nonumber \\
&\quad + i \hbar g \, \pc{\delta\hat{a}^\dagger \alpha + \delta\hat{a} \alpha^*} \pc{\hat{f}^\dagger + \hat{f}} - i \hbar \xi \pc{\alpha^* - \alpha}. 
\end{align}
The same linearisation process can be applied to the effective Lindblad operator
\begin{align}
\tilde{\mathcal{L}} \pc{\hat{O}} &= \frac{\gamma}{2}  \px{2 \delta\hat{a}^\dagger \hat{O} \delta\hat{a} - \hat{O}\delta\hat{a}^\dagger \delta\hat{a} - \delta\hat{a}^\dagger\delta \hat{a} \hat{O}} \nonumber \\
&\quad + \frac{\gamma}{2} \pc{ \alpha\pr{\delta\hat{a}^\dagger, \hat{O}}+ \alpha^* \pr{\hat{O}, \delta\hat{a}}}.
\end{align}
The value of $\alpha$ can be easily found via the equations of motion for both $\hat{a} \to \delta \hat{a} + \alpha$ and $\hat{f} \to \delta\hat{f} + \beta$ and by considering the fact that $\av{\delta\hat{ a}}=\av{\delta \hat{f}}=0$, such that $
 \alpha = \xi / \pc{i \omega + \gamma / 2}$ and $  \beta = -i g \md{\alpha}^2/\nu$.

\section{Analysis of the cooling process}

Having determined an effective Hamiltonian for the system, the cooling dynamics will be studied using the master equation
\begin{eqnarray}
\av{\dot{O}} &= -\frac{i}{\hbar} \av{\pr{\hat{O}, \hat{\tilde{H}}}}+ \frac{1}{2} \gamma
\av{2 \hat{a}^\dagger \hat{O} \hat{a} - \hat{O} \hat{a}^\dagger  \hat{a} - \hat{a}^\dagger \hat{a} \hat{O}  }
\nonumber \\
&+ \frac{1}{2} \gamma \av{ \alpha\pr{\hat{a}^\dagger, \hat{O}}+ \alpha^* \pr{\hat{O}, \hat{a}}}
. \nonumber 
\end{eqnarray}
Using the commutation rules and the total Hamiltonian we should apply this equation to the mean flexural mode number, the mean phonon number and the coherences:

\begin{align}
\hat{m} &= \av{\hat{f}^\dagger \hat{f}},\\
\hat{n} &= \av{\hat{a}^\dagger \hat{a}},\\
\hat{k}_1 &= i \pc{\hat{f}^\dagger + \hat{f}},  \\
\hat{k}_2 &= \hat{f}^\dagger - \hat{f},  \\
\hat{k}_3 &= \alpha \hat{a}^\dagger + \alpha^* \hat{a},  \\
\hat{k}_4 &= i \pc{\alpha \hat{a}^\dagger - \alpha^* \hat{a}},  \\
\hat{k}_5 &= i \pc{\alpha \hat{a}^\dagger + \alpha^* \hat{a}} \pc{\hat{f}^\dagger + \hat{f}},  \\
\hat{k}_6 &=  \pc{\alpha \hat{a}^\dagger + \alpha^* \hat{a}} \pc{\hat{f}^\dagger - \hat{f}} , \\
\hat{k}_7 &=  \pc{\alpha \hat{a}^\dagger - \alpha^* \hat{a}} \pc{\hat{f}^\dagger + \hat{f}} , \\
\hat{k}_8 &= i \pc{\alpha \hat{a}^\dagger - \alpha^* \hat{a}} \pc{\hat{f}^\dagger - \hat{f}} , \\
\hat{k}_9 &= \hat{f}^{\dagger2} + \hat{f}^2 , \\
\hat{k}_{10} &= i \pc{ \hat{f}^{\dagger 2} - \hat{f}^2} , \\
\hat{k}_{11} &= \pc{\alpha \hat{a}^{\dagger}}^{2} + \pc{\alpha^* \hat{a}}^2,  \\
\hat{k}_{12} &= i \pc{ \pc{\alpha \hat{a}^{\dagger}}^{2} - \pc{\alpha^* \hat{a}}}^2  
\end{align}
to obtain a closed set of differential equations. These are
\begin{align}
\dot{n} &= g \hat{k}_7 - \gamma \hat{n} -\frac{\gamma}{2} \hat{k}_3,\\
\dot{m} &= g \md{\alpha}^2 \hat{k}_2 + g \hat{k}_6,
\end{align}
and
\begin{align}
\dot{k}_1 &= - \nu \hat{k_2},\\
\dot{k}_2 &= \nu \hat{k}_1 -2 g \md{\alpha}^2 -2 g \hat{k}_3,\\
\dot{k}_3 &= \omega \hat{k}_4 - \gamma \md{\alpha}^2 -\frac{\gamma}{2}\hat{k}_3, \\
\dot{k}_4 &= - \omega \hat{k}_3 -2 g \md{\alpha}^2 \hat{k}_1 - \frac{\gamma}{2} \hat{k}_4,
\end{align}
while
\begin{align}
\dot{k}_5 &= - \nu \hat{k}_6 - \omega \hat{k}_7 -\frac{\gamma}{2} \hat{k}_5 - \gamma \md{\alpha}^2 \hat{k}_1,\\
\dot{k}_6 &= \nu \hat{k}_5 - 2 g \md{\alpha}^2 \hat{k}_3 -2 g \hat{k}_{11} -4 g\md{\alpha}^2 \hat{n} -2 g \md{\alpha}^2  \nonumber \\
&\quad + \omega \hat{k}_8 -\frac{\gamma}{2} \hat{k}_6 - \gamma \md{\alpha}^2 \hat{k}_2,
\\
\dot{k}_7 &= \nu \hat{k}_8 + \omega \hat{k}_5 -2 g \md{\alpha}^2\hat{k}_9 - 4 g \md{\alpha}^2 \hat{m} 
\nonumber \\
&\quad- 2 g \md{\alpha}^2 - \frac{\gamma}{2} \hat{k}_7,  \\
\dot{k}_8 &= - \nu \hat{k}_7 - 2 g \md{\alpha}^2  \hat{k}_4 - 2 g \hat{k}_{12} - \omega \hat{k}_6 \nonumber \\
&\quad -2 g \md{\alpha}^2 \hat{k}_{10} - \frac{\gamma}{2} \hat{k}_8, 
\end{align}
\begin{align}
\dot{k}_9 &= \nu \hat{k}_{10} - 2 g \md{\alpha}^2 \hat{k}_2 - 2 g \hat{k}_6,\\
\dot{k}_{10} &= - \nu \hat{k}_9 -2 g \md{\alpha}^2 \hat{k}_1 -2 g \hat{k}_5,\\
\dot{k}_{11} &= \omega \hat{k}_{12} -2 g \md{\alpha}^2 \hat{k}_7 - \gamma \hat{k}_{11} - \gamma \md{\alpha}^2 \hat{k}_3,\\
\dot{k}_{12} &= - \omega \hat{k}_{11} -2 g \md{\alpha}^2 \hat{k}_5 - \gamma \hat{k}_{12} - \gamma \md{\alpha}^2 \hat{k}_4.
\end{align}

\subsection{Stationary states}

In order to calculate the stationary flexural number $m_\text{SS}$ we set the right-hand side of the above cooling equations equal to zero, this yields,
\begin{align}
k_1^\text{SS} &= -\frac{2 \left(\gamma ^2 g \left| \alpha \right| ^2-4 g \omega ^2 \left| \alpha \right| ^2\right)}{16 g^2 \omega  \left| \alpha \right| ^2+\gamma ^2 \nu +4 \nu  \omega ^2}, \\
k_2^\text{SS} &= 0,\\
k_3^\text{SS} &= -\frac{2 \left(\gamma ^2 \nu  \left| \alpha \right| ^2+8 g^2 \omega  \left| \alpha \right| ^4\right)}{16 g^2 \omega  \left| \alpha \right| ^2+\gamma ^2 \nu +4 \nu  \omega ^2},\\
k_4^\text{SS} &= \frac{4 \gamma  \left(\nu  \omega  \left| \alpha \right| ^2+2 g^2 \left| \alpha \right| ^4\right)}{16 g^2 \omega  \left| \alpha \right| ^2+\gamma ^2 \nu +4 \nu  \omega ^2}
\end{align}

Defining the cubic frequencies $\mu^3 =\nu  \left(\gamma ^2+4 \omega ^2\right)+16  g^2 \omega \md{\alpha}^2 $  and $\lambda^3 = \nu  \left(\gamma^2 + \omega^2 \right) + 9  g^2 \omega \md{\alpha}^2$, we obtain
\begin{align}
k_5^\text{SS} &= \frac{g}{\lambda^3 \mu^3} \times \left(4 g^2 \omega  \left| \alpha \right| ^6 \left(5 \gamma ^2-16 \omega ^2\right) \right.\nonumber \\
&\quad \left. +2 \left| \alpha \right| ^4 \left(2 \gamma ^4 \nu +\gamma ^2 \omega  \left(8 g^2-7 \nu  \omega \right)+8 g^2 \omega ^3\right)+ \right.\nonumber \\
&\quad \left.\md{\alpha}^2  \nu  \left(\gamma ^2+\omega ^2\right) \left(\gamma ^2+4 \omega ^2\right) \right),
\end{align}
\begin{align}
k_6^\text{SS} &= 0,\\
k_7^\text{SS} &= \frac{\gamma  g }{2 \omega  \lambda^3 \mu^3} \times \left(16 g^2 \omega  \left| \alpha \right| ^6 \left(\gamma ^2-5 \omega ^2\right)
\right.\nonumber \\
&\quad \left.
-2 \omega  \left| \alpha \right| ^4 \left(\nu  \omega  \left(8 \omega ^2-\gamma ^2\right)+8 g^2 \left(\gamma ^2+\omega ^2\right)\right)
\right.\nonumber \\
&\quad \left.
- \md{\alpha}^2  \nu  \left(\gamma ^2+\omega ^2\right) \left(\gamma ^2+4 \omega ^2\right) \right),\\
k_8^\text{SS} &= \frac{g}{\omega  \lambda^3 \mu^3} \times
 \left(96 g^4 \omega ^2 \left| \alpha \right| ^8
\right.\nonumber \\
&\quad \left.
 +2 \nu  \omega  \left| \alpha \right| ^4 \left(3 \gamma ^2 \nu  \omega +2 g^2 \left(7 \gamma ^2
 +16 \omega ^2\right)\right)
 \right.\nonumber \\
&\quad \left.+16 g^2 \omega  \left| \alpha \right| ^6 \left(\gamma ^2 \nu +2 \omega  \left(6 g^2+\nu  \omega \right)\right)\right.\nonumber \\
&\quad \left.+ \md{\alpha}^2  \nu ^2 \left(\gamma ^2+\omega ^2\right) \left(\gamma ^2+4 \omega ^2\right)\right)
,\\
k_9^\text{SS} &= -\frac{2 g^2 }{\nu   \lambda^3 \mu^3} \times \left(2 g^2 \omega  \left| \alpha \right| ^6 \left(\gamma ^2+4 \omega ^2\right)
\right.\nonumber \\
&\quad \left.
+2 \left| \alpha \right| ^4 \left(\gamma ^4 \nu -4 \gamma ^2 \omega  \left(\nu  \omega -2 g^2\right)
+4 \omega ^3 \left(2 g^2+\nu  \omega \right)\right)
\right.\nonumber \\
&\quad \left.+ \md{\alpha}^2  \nu  \left(\gamma ^2+\omega ^2\right) \left(\gamma ^2+4 \omega ^2\right) \right),\\
k_{10}^\text{SS} &= 0,\\
k_{11}^\text{SS} &= \frac{\left| \alpha \right| ^4 }{\omega \lambda^3 \mu^3} \times \left(-16 g^4 \omega  \left| \alpha \right| ^4 \left(\gamma ^2-8 \omega ^2\right)
\right.\nonumber \\
&\quad \left.
+16 \md{\alpha}^2  g^2 \omega \left(\gamma ^2 \nu  \omega +g^2 \left(\gamma ^2-2 \omega ^2\right)\right)\right)\nonumber \\
&\quad + \frac{\nu  \left(\gamma ^2-2 \omega ^2\right) \left(2 \gamma ^2 \nu  \omega +g^2 \left(\gamma ^2+4 \omega ^2\right)\right)}{\omega \lambda^3 \mu^3}, \\
k_{12}^\text{SS} &= -\frac{\gamma  \left| \alpha \right| ^4 }{\lambda^3 \mu^3} \times\left(96 g^4 \omega  \left| \alpha \right| ^4\right.\nonumber \\
&\quad \left.
+16 \md{\alpha}^2  g^2\left(\gamma ^2 \nu +\omega  \left(3 g^2+2 \nu  \omega \right)\right)\right.\nonumber \\
&\quad \left.+3 \nu  \left(2 \gamma ^2 \nu  \omega +g^2 \left(\gamma ^2+4 \omega ^2\right)\right)\right)
\end{align}
\begin{widetext}
and
\begin{align}
n_\text{SS} &=-\frac{-2 \gamma ^4 \nu ^2 \omega  \left| \alpha \right| ^2-2 \gamma ^2 \nu ^2 \omega ^3 \left| \alpha \right| ^2-16 \gamma ^2 g^4 \omega  \left| \alpha \right| ^6+16 \gamma ^2 g^4 \omega  \left| \alpha \right| ^4-64 g^4 \omega ^3 \left| \alpha \right| ^6}{2 \omega \lambda^3 \mu^3} \nonumber \\
&\quad-\frac{16 g^4 \omega ^3 \left| \alpha \right| ^4+\gamma ^4 g^2 \nu  \left| \alpha \right| ^2-36 \gamma ^2 g^2 \nu  \omega ^2 \left| \alpha \right| ^4+5 \gamma ^2 g^2 \nu  \omega ^2 \left| \alpha \right| ^2+4 g^2 \nu  \omega ^4 \left| \alpha \right| ^2}{2 \omega \lambda^3 \mu^3}, \\
m_\text{SS} &= -\frac{\left(\gamma ^2+\omega ^2\right) \left(-48 g^2 \omega  \left| \alpha \right| ^2-3 \gamma ^2 \nu +4 \nu ^3-12 \nu  \omega ^2\right)}{48 \nu  \omega  \lambda^3} \nonumber\\
&\quad +\frac{\left| \alpha \right| ^2 \left(16 \gamma ^2 g^2 \omega  \left| \alpha \right| ^2+48 g^2 \nu ^2 \omega  \left| \alpha \right| ^2-32 g^2 \omega ^3 \left| \alpha \right| ^2+\gamma ^4 \nu +4 \gamma ^2 \nu ^3+2 \gamma ^2 \nu  \omega ^2+8 \nu ^3 \omega ^2-8 \nu  \omega ^4\right)}{32 \nu  \omega  \lambda^3} \nonumber \\
&\quad +\frac{\left(\gamma ^2 g \left| \alpha \right| ^2-4 g \omega ^2 \left| \alpha \right| ^2\right) \left(-192 g^4 \omega ^2 \left| \alpha \right| ^4+16 g^2 \nu ^3 \omega  \left| \alpha \right| ^2-\gamma ^4 \nu ^2-4 \gamma ^2 \nu ^4+18 \gamma ^2 \nu ^2 \omega ^2+8 \nu ^4 \omega ^2-8 \nu ^2 \omega ^4\right)}{32 g \nu  \omega  \lambda^3 \mu^3}
\nonumber \\
&\quad +\frac{3 \omega -2 \nu }{6 \omega }.\label{eq:mSS}
\end{align}

\subsection{Cooling dynamics}

To calculate the effective cooling rate of the flexural modes $\gamma_\text{eff}$, we notice that the mean flexural mode number $\hat{m}$ evolves on a relatively slow time scale comparative to all other variables. These can be eliminated adiabatically leaving us with only a single effective cooling equation. Setting all time derivatives equal to zero except the one evolving the flexural modes, we find the expressions for $\hat{k}_2 = 0 $ and $\hat{k}_6 = \mathcal{F}(m(t))$, i.e., $\hat{k}_6 $ is a function of the mean flexural number. Substituting these in $\dot{\hat{m}} (t) = g \md{\alpha}^2 \hat{k}_2 + g \hat{k}_6$, we obtain an expression where $\dot{\hat{m}} (t) = g \mathcal{F} (m(t))$. Solving this equation yields finding $m(t) = a e^{\gamma_{\text{eff}} t} + m_\text{SS}$ 
with $m_\text{SS}$ given by Eq.~\eqref{eq:mSS}, 
\begin{align}
a &= \left\lbrace \frac{1}{\nu  \omega } \times \left[32 g^4 \omega ^2 \left| \alpha \right| ^8 \left(\gamma ^2+4 \left(3 \nu ^2+\omega ^2\right)\right) \right. \right. \nonumber \\
&\quad +16 g^2 \omega  \left| \alpha \right| ^6 \left(\gamma ^4 \nu +2 \gamma ^2 \left(8 g^2 \omega +2 \nu ^3+\nu  \omega ^2\right)-8 \omega  \left(g^2 \left(-6 \nu ^2+9 \nu  (2 m(0) +1) \omega -2 \omega ^2\right)-\nu ^3 \omega +\nu  \omega ^3\right)\right)\nonumber \\
&\quad + \md{\alpha}^2  \nu ^2 \left(\gamma ^2+\omega ^2\right) \left(\gamma ^2+4 \omega ^2\right) \left(\gamma ^2+4 \left(\nu ^2-2 \nu  (2 m(0)+1) \omega +\omega ^2\right)\right) \nonumber \\
&\quad + 2 \nu  \omega  \left| \alpha \right| ^4 \left[ \gamma ^2 \nu  \omega  \left(7 \gamma ^2+12 \nu ^2-20 \omega ^2\right) \right. \\ \nonumber
&\quad \left. \left. \left. +4 g^2 \left(4 \gamma ^4+\gamma ^2 \left(14 \nu ^2-25 (2 m(0) +1) \nu  \omega +20 \omega ^2\right)+4 \omega ^2 \left(8 \nu ^2-13 (2 m (0) +1) \nu  \omega +4 \omega ^2\right)\right) \right] \right] \right\rbrace \\ \nonumber
&\quad \left/ \px{ 16 \left(g^2 \omega  \left| \alpha \right| ^4 \left(25 \gamma ^2 \nu +144 \md{\alpha}^2  g^2 \omega+52 \nu  \omega ^2\right)+ \md{\alpha}^2  \nu ^2 \left(\gamma ^2+\omega ^2\right) \left(\gamma ^2+4 \omega ^2\right) \right)\right.}
\end{align}
where $m(0)= \pc{\exp \left(\frac{\hbar \nu}{k_B T_{\text{graph}}} \right)-1}^{-1}$ is the occupation number calculated at $t=0$ and initial temperature $T_\text{graph}$, and cooling rate given by
\begin{align}
\gamma_\text{eff} &=\px{64 \gamma  g^2 \lambda ^3 \nu ^2   \omega  \left| \alpha \right| ^2} \nonumber \\
&\quad \Big/ \left\lbrace 32 g^2 \omega  \left| \alpha \right| ^2 \left(8 g^2 \omega  \left| \alpha \right| ^2 \left(\gamma ^2-3 \nu ^2+\omega ^2\right)+\nu  \left(\gamma ^4+5 \omega ^2 \left(\gamma ^2-2 \nu ^2\right)-\gamma ^2 \nu ^2+4 \nu ^4+4 \omega ^4\right)\right) \right. \nonumber\\
&\quad \left. +\nu ^2 \left(\gamma ^2+\omega ^2\right) \left(\gamma ^4+8 \gamma ^2 \left(\nu ^2+\omega ^2\right)+16 \left(\nu ^2-\omega ^2\right)^2\right) \right\rbrace.
\end{align}

\end{widetext}

\end{document}